\documentclass[12pt]{article}
\newcommand{\be}{\begin{equation}}
\newcommand{\ee}{\end{equation}}
\newcommand{\ba}{\left( \begin{array}}
\newcommand{\ea}{\end{array} \right)}

\begin{document}

\title{Crossover between displacive and order-disorder phase transition}

\author{A.N. Rubtsov,$^a$ J. Hlinka,$^b$ T. Janssen$^c$}

\maketitle

\begin{center}
$a$ Physics Department, Moscow State University,\\ Moscow, Russia\\
$b$ Institute of Physics, Czech Academy of Science,\\ Prague, Czech Republic,\\
$c$ Institute of Theoretical Physics, University of Nijmegen,\\ Nijmegen, The
Netherlands.
\end{center}

submitted to Phys. Rev. E

\begin{abstract}

{The phase transition in a 3D array of classical anharmonic
oscillators with harmonic nearest-neighbour  coupling (discrete
$\phi^4$ model) is studied by Monte Carlo (MC) simulations and by
analytical methods. The model allows to choose a single
dimensionless parameter $a$ determining completely the behaviour
of the system. Changing $a$ from 0 to $+\infty$ allows to go
continuously from the displacive to the order-disorder limit.
We calculate   the  transition temperature $T_c$
and the temperature dependence of the order parameter down to
$T=0$ for a wide range of the parameter $a$. The  $T_c$ from MC
calculations shows an excellent agreement with the known
asymptotic values for small and large $a$. The obtained MC results
are further compared with predictions of the  mean-field and
independent-mode approximations as well as with predictions
of our own approximation scheme.

In this approximation, we introduce an auxiliary system, which
yields approximately the same temperature behaviour of the order
parameter, but allows the decoupling of the phonon modes.

Our approximation gives the value of $T_c$ within an error of
$5\%$ and satisfactorily describes the temperature dependence of the
order parameter for all values of $a$.}
\end{abstract}

\thispagestyle{empty}

\section{INTRODUCTION}

One of the basic classification schemes for  structural phase
transitions consists in assigning it to the order-disorder or the
displacive type. The displacive transition can be described as a
freezing of a phonon mode, which shows "critical softening" at the phase
transition point.
The occurrence of a soft mode is often used as
criterion for a displacive transition in a real systems, since the
frequency of the phonon modes is accessible by spectroscopic
experiments.

In the order-disorder case, there are two or more locations
for each atom in the unit cell. Occupation numbers for
these locations are the same above the transition temperature, and
differ below. Formally, as in the displacive case, the system can be
 described in  "phonon" language.

There is a simple model which shows that one can
go from the order-disorder to the displacive type
{\it continuously} \cite{Aubry}. This model can be defined as a 3D
cubic lattice of classical anharmonic 1D oscillators with
nearest-neighbour harmonic coupling \cite{Cowley,SS,Radescu,Padle,Giddy,Heine}:

\begin{equation}
V=\frac{A}{2} \sum_{n} x_{n}^2  + \frac{B}{4} \sum_{n} x_{n}^4  +
\frac{C}{2} \sum_{n, n'} (x_{n}-x_{n'})^2 \sigma(n, n'),
\label{E1}
\end{equation}
$A, B$, and $C$ are model parameters, the indices $n$ and $n'$ run
over all oscillators, $\sigma(n,n')$ is equal to 1 for neighbouring
particles and vanishes elsewhere. The system undergoes a phase
transition from the higher symmetry to the lower symmetry phase at a
certain temperature $T_c$ for any $A<0, B>0, C>0$, {\it i.e.} the
statistical average of each  coordinate $x_n$ takes a non-zero
value $\eta =<x_n>$ below $T_c$ and vanishes above. It is often
convenient to express the potential (1) as
\begin{equation}
V= \sum_{n} v(x_n)  - C \sum_{n, n'}  x_{n}x_{n'} \sigma(n, n'),
\label{E2prime}
\end{equation}
with an "on-site" single particle potential
\begin{equation}
v(x)=\frac{A'}{2} x^2  + \frac{B}{4} x^4,~~ A'=A+12C~. \label{sitep}
\end{equation}
It is known that the behaviour of the system is governed by the
ratio
\begin{equation}
a=-A/C.
\end{equation}

At small $a>0$ the system shows a displacive phase transition,
while for large $a$ the system behaves as the Ising model, which
shows a typical order-disorder phase transition. The transition
temperatures $T_c$ in the limit cases are known from Ising-model
and self-consistent phonon calculations, to be respectively
\cite{Cowley,Radescu,Bruce,Opper}:

\begin{eqnarray}
T_c(a\downarrow 0) \approx 2.64 C|A|/B, \nonumber \\ T_c(a\rightarrow +
\infty) \approx
9.12 C|A|/B ,   \label{E_lim}
\end{eqnarray}
assuming here and further that the temperature is expressed in
energy units (the Boltzman constant  equal to 1). On the other
hand, despite of the important role  of the above model in the
theory of the structural phase transitions \cite{Cowley}, the
actual dependence of $T_c(a)$ is not known. The results of previous
molecular dynamics and Monte Carlo studies are collected in Figure 1.
They obviously do not give a consistent quantitative picture.
So far the analytical study was restricted to the mean-field approach.
\cite{Cowley,JaTj}.

Recently, it was observed that the knowledge of the dependence
$T_c(a)$ can be useful in the quantitative analysis  of the
properties of crystalline  $Sn_2P_2S_6$  which has a ferroelectric
phase transition showing simultaneously features typical for both
the order-disorder and  displacive type  \cite{Hlinka}.

The aim of this paper is to  establish this dependence of $T_c(a)$
as well as the temperature behaviour of the order parameter. Let us
stress that, similarly to some related
papers \cite{Radescu,Padle,Giddy, Heine}, we are not interested
here in details of critical behaviour  in the very vicinity of the
phase transition. Critical behaviour of this model is thoroughly
described  for example in the reference \cite{Cowley}.

The paper is organized as follows. Section 2 describes our MC
simulations performed for a wide range of values of the parameter $a$.
In section 3, we first compare the MC results with rather poor
predictions of the standard decoupling schemes
and suggest an improved self-consistent equation for the order parameter
that allows to calculate
both the transition temperature and the order parameter with a
reasonable accuracy for all values of $a$.

\section {MONTE-CARLO SIMULATIONS}

For numerical simulation it is convenient to re-scale coordinates
and energy units. This allows to reduce the
potential energy (\ref{E1}) into the form
\begin{equation}
V_{\rm red}=-\frac{a}{2} \sum_{n} x_{n}^2  + \frac{a}{4} \sum_{n}
x_{n}^4  + \frac{1}{2} \sum_{n, n'} (x_{n}-x_{n'})^2 \sigma(n,
n'), \label{E2}
\end{equation}
with a single dimensionless parameter  $a=-A/C$. Then the
re-scaled order parameter at zero temperature is equal to 1 for
any $a>0$.

The typical size of the array of atoms studied in our MC
simulations is $10 \times 10 \times 10$ atoms, with periodic
boundary conditions.
We perform Monte-Carlo steps consecutively for each atom, and
accept (or reject) them accordingly to standard criteria.
Additionally, we perform "magic" steps
for the case of large $a$, when the sign of the coordinate of the
given atom may flip. These steps allow the system to thermalize in
the order-disorder limit as well. We calculate the square of the order
parameter as the average
\begin{equation}
\eta^2=N^{-1/2} <X_0^2-X_1^2>,  \label{num}
\end{equation}
where $X_k = N^{-1/2} \sum x_n e^{i k n}$ is the Fourier transform of
$x_n$,
$N$ is the total number of particles.

For the case of an infinite slab, $N^{-1/2} X_1^2$ is negligible, and
(\ref{num})
gives purely the square of the order parameter. For a finite
slab, the term $X_1^2$ allows to remove fluctuations from the
high-temperature branch.

The results of the calculations are presented in Figures 2-4.
It is crucial to check the dependence of the results on the
system size. Figure 2 presents the temperature dependence of $X$
for sizes  $15 \times 15 \times 15$ and  $5 \times 5 \times 5$. It
is clear, that change of the size of the slab affects practically
only the fluctuation region near $T_c$. The value of $T_c$
calculated from the fit of the dependence (see Figure 2) remains
almost unchanged. This type of  size dependence of the data is
found for the whole range of $a$.

Figure 3 presents data for $\eta^2(T)$ obtained for the potential
(\ref{E2}) for   different values of the parameter $a$. Note that
the Landau theory yields a linear temperature dependence
for $\eta^2(T)$ .

Values of $T_c$ are extracted from the data presented in Figure 2.
The plot for $T_c(a)$ is given in Figure 4 where a logarithmic
scale for the $a$-axis is used. The monotonic dependence
approaches known limit values with a good accuracy. The change in
$a$ is for which $T_c(a)$ varies significantly is about 2 orders of
magnitude.




\section {ANALYTICAL APPROACHES}

Two standard decoupling schemes  have been used in the literature to make the
phase transition in the model tractable, usually referred to as mean
field (or independent site) approximation and self-consistent
phonon (or independent mode) approximation. In this section we
first analyse the advantages and disadvantages of these standard
approximations and then we propose a modified approximation scheme
that combines advantages of both schemes.

\subsection {Independent mode approximation }
In the independent mode approximation (IMA), the deviations from
the average value given by the  order parameter
\begin {equation}
y_n= x_n- \eta
\end {equation}
are represented by  Fourier coordinates $Y_k=N^{-1/2} \sum_n y_n
e^{i k n}$. Interaction between Fourier coordinates is simplified
by assuming that each Fourier coordinate is influenced only by
the average of its interactions with the other coordinates. This
leads to an effective harmonic approximation. The order parameter
in IMA is defined by the equation \cite{Cowley}
\begin {equation}
A \eta + B \eta^3+3 B \eta I(T)=0,
\label{EW cond}
\end{equation}
where the function $I(T)=N^{-1} \sum_k Y_k Y_{-k}$ is
calculated from the phonon dispersion relation
renormalized by the given value of the  order parameter and the thermal
fluctuations. In the vicinity of the phase transition point,
$I(T)$ can be evaluated by assuming a "critical" phonon dispersion
(with  zero frequency of the zone-center mode):
\begin {equation}
I(T) \approx \frac {T}{4C(2 \pi)^3} \int{\frac{d^3
k}{3-\cos k_x-\cos k_y-\cos k_z}}=\frac {T}{3 C \kappa}
\label{IMA}
\end {equation}
where $\kappa \approx 2.638~$. Note that  the stability
limit $T_{\rm c, IMA}=-AC \kappa /B$ of the high temperature  phase as
obtained from (\ref{EW cond}) and (\ref{IMA}), provides an exact
prediction for $T_c$ and $\eta(T)$  in the displacive
limit.  However, for the order-disorder limit, the IMA
values differ considerably from the exact values.

\subsection {Mean field approximation }

In the mean-field approximation (MFA) for the system with
a harmonic coupling, the behaviour of the
original system is modelled by an auxiliary system in which all
direct inter-site interactions are replaced by an
effective external field $E$, but the on-site anharmonicity is
kept without any approximation. Taking the  on-site potential as
given by (\ref{sitep}), the ensemble averages in such an auxiliary system at
fixed external field are given by
\begin{equation}
\eta=<x_n> = g_T (E) \equiv \frac{\int x \exp [-(v(x)-Ex)/T] d
x}{\int \exp [-(v(x)-Ex)/T] d x}. \label{defF}
\end{equation}
Since at finite temperatures the  $g_T(E)$ is a  monotonic
function,
it can be inverted and  the self-consistent equation for the order
parameter in the auxiliary system can be written as
\begin{equation}
E=g_T^{-1}(\eta). \label{MF eq}
\end{equation}
The effective field $E$ is defined as the force on $x_n$ supplied
by the interaction terms separated in (\ref{E2prime}), assuming that the
displacement of the six nearest neighbouring sites (or at least
their sum) is frozen at the equilibrium value $\eta$ :
\begin{equation}
E= 12 C \eta . \label{MF eq1}
\end{equation}
Self-consistent solution of equations (\ref{MF eq},\ref{MF eq1})
defines the order parameter $\eta_{\rm MFA}$ in MFA.
The phase transition temperature  $T_{\rm c, MFA}(a)$
at which $\eta_{\rm MFA}$ vanishes is shown in Figure 1. It was
previously remarked by  S. Aubry \cite{Aubry,Cowley} that the
relative over-estimation of $T_c$ by  MFA is almost the same (about 30
percent) for both limit cases ($a \rightarrow +0 , a \rightarrow
+\infty $). Comparison of $T_{\rm c, MFA}(a)$ with our MC results
shows that the discrepancy is really systematic for all
intermediate cases. Although this error is rather large, its
systematic character  strongly suggests that the physics of the
crossover is already well taken into account by the MFA.

Let us analyse the function $g_T(E)$
describing the auxiliary ensemble of the uncoupled on-site
oscillators in more detail. Let us stress the following points:
\begin{enumerate}
\item
The variation of $T_c$ with $a$ is within MFA entirely given by
the slope of the  function $g_T(\eta)$ at $\eta=0$.
\item
Unlike the on-site potential $v(x)$, the function $g_T(E)$ at
finite temperature is a smooth monotonic odd function
(see Figure 5) at any $T, a$ \cite{JaTj}.
Both $g_T(E)$ and its inverse  $g_T^{-1}(\eta )$ can
be expanded in Taylor series:
\begin{equation}
g_T(E) =\sum_{i=1}^{\infty} \chi_{2i-1}(T) E^{2i-1},~~
g_T^{-1}(\eta) =\sum_{i=1}^{\infty} \xi_{2i-1}(T) \eta^{2i-1}
\label{T_exp}
\end{equation}
\item The function  $g_T^{-1}(\eta)$
can be identified with the derivative of its free
energy $F(\eta,T)$, which can  thus be written in the form
\begin{equation}
F(\eta,T) = F(0,T) +\sum_{i=1}^{\infty}
\frac{\xi_{2i-1}(T)}{(2i)} \eta^{2i}
\end{equation}
\item
Obviously, the Taylor expansion coefficients of $g_T(E)$ and
$g_T^{-1}(\eta)$ are related ($\xi_1 \chi_1 =1$, $ \xi_3 \chi_1^3
+  \chi_1 \xi_3^3 =0$, etc.) This allows to express
$\xi_{2i-1}(T)$ in the limit case of the weak anharmonicity ($B<<
A'$) by expanding $g_T(E)$ in powers of $B$. With an accuracy $O(B^2)$
we obtain
\begin{equation}
\xi_1 (T) = A' + \frac{3BT}{A'} ~ , ~~\xi_3 (T)= B~. \label{OB}
\end{equation}
In the strongly anharmonic order-disorder limit ($
A'<0, T<< A'^2/B$), expressing $g_T(E)$ via  averages $<x^2>,
<x^4>$ yields
\begin{equation}
\xi_1 (T) = \frac{BT}{A'} ~ , ~~\xi_3 (T) =
  \frac{B^2  }{3A'^2}~, ~~~...~~. \label{lim2}
\end{equation}
\item Finally, let us note that in the weak anharmonicity case
we can solve the inverse problem - express the parameters of the
on-site potential via the first two free energy coefficients
 $\xi_1(T),\xi_3(T) $. With the same accuracy as (\ref{OB})
\begin{equation}
A' = \xi_1 (T)  -\frac{3 \xi_3(T) T}{\xi_1 (T) } ~ , ~~B =
\xi_3 (T)~~. \label{inve}
 \end{equation}
\end{enumerate}

\subsection {Combined scheme}
We have seen that the IMA predicts well the phase transition
temperature in the displacive limit, while MFA predicts rather
well its variation with $a$. It would be desirable to have an
approximate equation of state for the system (\ref{E1}) that combines the
advantages of both above discussed approaches. The key idea of our
approach is the assumption of the existence of  an effective potential
(with temperature dependent coefficients) for which the
self-consistent phonon approximation gives correctly the order parameter.
In determining the coefficients of such an effective
potential, we use the properties of the free energy $F(\eta ,T)$ (respectively
its derivative
$g_T^{-1}(\eta)$) of  the auxiliary system of uncoupled anharmonic
oscillators discussed above.

More precisely, the self-consistent equation for $\eta(T)$ is
constructed in three steps, as follows:

\begin{enumerate}
\item We look for an effective  on-site potential of the form
\begin{equation}
u(x)=\frac{\alpha'}{2} x^2  + \frac{\beta}{4}  x^4 \label{softpot}
\end{equation}
where $\alpha'$ and $\beta$ are defined  by the  expressions
that appear in the above discussed inverse problem (\ref{inve}):
\begin{equation}
\alpha' = \xi_1   -\frac{3 \xi_3 T}{\xi_1 } ~ , ~~\beta =\xi_3. \label{inve2}
\end{equation}
This potential obviously coincides with $v(x)$ in the weak anharmonic
limit.
\item We introduce a function $\xi_{1,\rm eff}(T,\eta)$,
which allows to write $g_T^{-1}$ formally as a finite polynomial:
\begin{equation}
g_T^{-1}(\eta)\equiv \xi_{1,\rm eff}(T,\eta) \eta + \xi_3(T)
\eta^3. \label{xi1efe}
\end{equation}
These functions $\xi_{i,{\rm eff}} (T)$ are used instead of $\xi_1(T)$ in the
definitions
(\ref{inve2}), so that we have
\begin{equation}
\alpha' = \xi_{1,\rm eff}(T,\eta)   -\frac{3 \xi_3(T) T}
{\xi_{1,\rm eff}(T,\eta) } ~ , ~~\beta =  \xi_3(T).
\label{sub2}
\end{equation}
Note that the potential $u(x)$ still coincides with $v(x)$ in the
weak anharmonic limit for small $\eta$, since $\xi_{1,\rm
eff}(T,\eta)$ goes to $\xi_{1}(T)$ for $\eta \rightarrow 0$.
\item  We consider eq. (\ref{E2prime}) and replace $v(x)$ with coefficients
$A'$ and $B$  by an expression $u(x)$ with
temperature-dependent coefficients $\alpha '$ and $\beta$  defined in eq.
(\ref{sub2}). Then we apply IMA
to this auxiliary system. The eq. (\ref{EW cond}) then becomes
\begin{equation}
\left[   \xi_{1,\rm eff}(T,\eta) - 12 C + \frac{\xi_{3}(T)T}{C
\kappa} -\frac{3 \xi_{3}(T)T}{\xi_{1,\rm eff}(T,\eta)}    \right]
\eta +\xi_{3}(T)\eta^3 =0. \label{finale}
\end{equation}

This equation  is to be solved together with formulae
(\ref{xi1efe}) and (\ref{defF}), defining   $\xi_{1,\rm eff}$
and $g_T$, respectively.
The value of $\xi_3(T)$, entering these equations, is given by the
series (\ref{T_exp}).

\end{enumerate}

For the calculation of the phase transition
temperature only, the second step can be omitted. It is obvious
from its construction that the suggested method provides the same
(exact) result for the $T_c$ in the displacive limit as the usual
IMA. In the extreme order-disorder limit, the value of $T_c$
defined by (\ref{finale}) can be obtained analytically using
(\ref{lim2}). The resulting value of $T_c $
overestimates the known Ising value by less then 7\%. The
principal advantage of the modified approach is that it allows to
calculate the  $T_c$ (and $\eta(T)$) with the above or better accuracy
for
all values of $a$, as it can be seen from the comparison with our
MC data (Figure 6). The MC result for $\eta^2 (T)$ is satisfactorily
described as well (Figure 7).

\section {DISCUSSION}
Let us analyze the proposed model in
comparison with the standard decoupling schemes.
The latter treat the system as a gas of elementary
excitations, which are supposed to interact weakly.
The assumption of weak interaction allows to replace
the interaction between the elementary excitations with an interaction
with an average field.
Choice of the elementary excitations as the plane waves or on-site
oscillators yields IMA or MFA, respectively.
It is clear, however, that the assumed weakness of the interaction
is actually not realized for the general case, no matter what elementary
excitations we choose.

The main advantage of our approach is that it
virtually replaces the real strongly-correlated system (\ref{E1}) with
an auxiliary one, which allows  decoupling.
It is also worth noting that the theory is carried out
in terms of $g_T(\eta)$, which is always a smooth monotonic function.
Moreover, $g_T(\eta)$ does not change drastically when $a$ is varied
from $0$ to $+\infty$ - the calculation of parameters $\xi_{1}$ and $\xi_3$ at
$T_c$ shows  that their dimensionless values lie within the relatively narrow
ranges $9.7 ... 12$ and $0 ...
4$, respectively.
Therefore the replacement procedure works uniformly well for all values
of the parameters.

It would be interesting to investigate possible extensions to higher-order
terms.
A systematic extension of the present method should contain a
larger number of
terms in the effective on-site potential (\ref{softpot}) and
in the expression for the function $g_T^{-1}(\eta)$ in (\ref{sub2}), and
solve the self-consistent equation for the auxiliary system
more accurately than (\ref{finale}).

As a simplification we can consider
a purely linear auxiliary system, {\it i.e.}  neglect
the $\beta$ term in (\ref{softpot}) and
$\xi_3(T)$ in (\ref{sub2}). We obtain $g_T^{-1}(\eta) \equiv
\xi_{1,\rm eff}(T,\eta) \eta$ and (\ref{finale}) then reduces
simply to the mean-field equation of state $g_T^{-1}(\eta) = 12C \eta $.
Therefore, the scheme proposed here
can also be considered as
a generalization of the mean field approximation.

Our method can be applied to more complicated models
for  which the self-consistent phonon theory is exact in the weak
anharmonic limit. This is  particularly interesting for the
analysis of the DIFFOUR model \cite{Ted} in which the additional
second neighbour harmonic coupling shows a phase transition to an
incommensurate phase for which the MC calculations are much more
difficult.

\section{CONCLUSIONS}
We have studied the crossover from a displacive to an order-disorder
phase transition in the discrete $\phi^4$ model with first-neighbour
coupling.. The crossover is
governed by the single parameter $a$.
Quantitative information about $T_c(a)$ and $\eta(T,a)$ in this
simple model may be helpful in elucidating the behaviour of
some real crystals with phase transitions of a mixed displacive
and order-disorder type.

In terms of the dimensionless parameter $a$ we determined the
change of the transition temperature by Monte-Carlo calculations.
These show a crossover from the displacive to the order-disorder
limit.

Monte Carlo calculations have shown an excellent agreement for
 $T_c$ in the two limit cases in which exact results are known. We
  expect that the same precision is obtained for the intermediate
region. Thus, the presented Monte Carlo results can be taken as
quite reliable estimates of $T_c(a)$ with a precision of the  order of
1\% and we believe that a comparable precision  was achieved for
the temperature dependence of the order parameter (except in the
critical region in the vicinity of the phase transition).

We have presented an analytical approach, which goes beyond
the conventional decoupling schemes. For this, we introduce the
auxiliary
array of oscillators that (i) can be treated in the independent-mode
approximation and (ii) yields approximately the same values of $T_c$ and
the order parameter, as the real system.
The method  combines the equation of state of the
self-consistent phonon theory with the response function of the
system of uncoupled  anharmonic oscillators used in the the
mean-field theory. It can be presented as a generalization of the
mean-field scheme.

The analytical results for $T_c$ agree
with Monte-Carlo simulations with about 5$\%$ accuracy. Further
improvement could possibly come from
higher order terms in the expansion we have used. The
formalism can be used to study incommensurate phase transitions as
well.

\section{ACKNOWLEDGEMENT}

The work was supported by an INCO Copernicus program (Grant ERBICI 15 CT
970712).

\newpage

\begin{center} {\bf Figure captions}
\end{center}

Fig.1 The critical temperature vs. $\ln a$.
The curve shows the mean-field result; thin horizontal lines show the
asymptotic values of $T_c$.
Results of previous Molecular-Dynamics \cite{Padle,SS}
and Monte-Carlo \cite{Radescu} calculations are plotted with filled and
open
circles,
respectively.

Fig.2 The role of the finite size of the system studied numerically.
Numerical data for the square of the order parameter plotted as a
function of $T$.
Filled circles: $5\times 5\times 5$ oscillators; open circles: $15\times
15\times 15$ oscillators.
Both results are obtained for $a=5$ in  eq. (\ref{num}) by
averaging over $3000$ realizations at each point.
The solid line shows the interpolation used to obtain $T_c$.

Fig.3 The temperature dependence of the square of the order
parameter for values of $a$ varying from 0.98 to 4000. There is a factor
$\sqrt{2}$ between the $a$ values for the neighbouring curves.
The data are obtained using eq. (\ref{num}) by  averaging over
$1000$ realizations at each point; the relative amount of "magic"
steps is 0.02.

Fig.4 Numerical results for the critical temperature $T_c$ {\it vs.} $\ln a$.
The values of $T_c$
are extracted from the data presented in Fig.3.
Thin horizontal lines show the asymptotic values of $T_c$.

Fig.5 Typical dependence of $g_T^{-1}(\eta)$ at small $a$ (solid line)
and large $a$ (dashed line).The inset shows the on-site
potential for both cases. (Calculated for (\ref{E2})with $T=5$ and
$a= 1$ and $ 100$, respectively.)

Fig.6 Numerical data for $T_c(\ln a)$ compared with results of calculations by
equations
(\ref{finale},\ref{defF},\ref{xi1efe}) (the solid line). The mean-field
approximation is given by the dashed line.

Fig.7 The temperature dependence of the square of the order parameter at
several
values of $a (=0.98, 3.9, 15.6, 62.5, 250, 1000, 4000)$. $T_c$ grows with
increasing  $a$: numerical data (points) and calculation from
(\ref{finale},\ref{defF},\ref{xi1efe}) (lines).

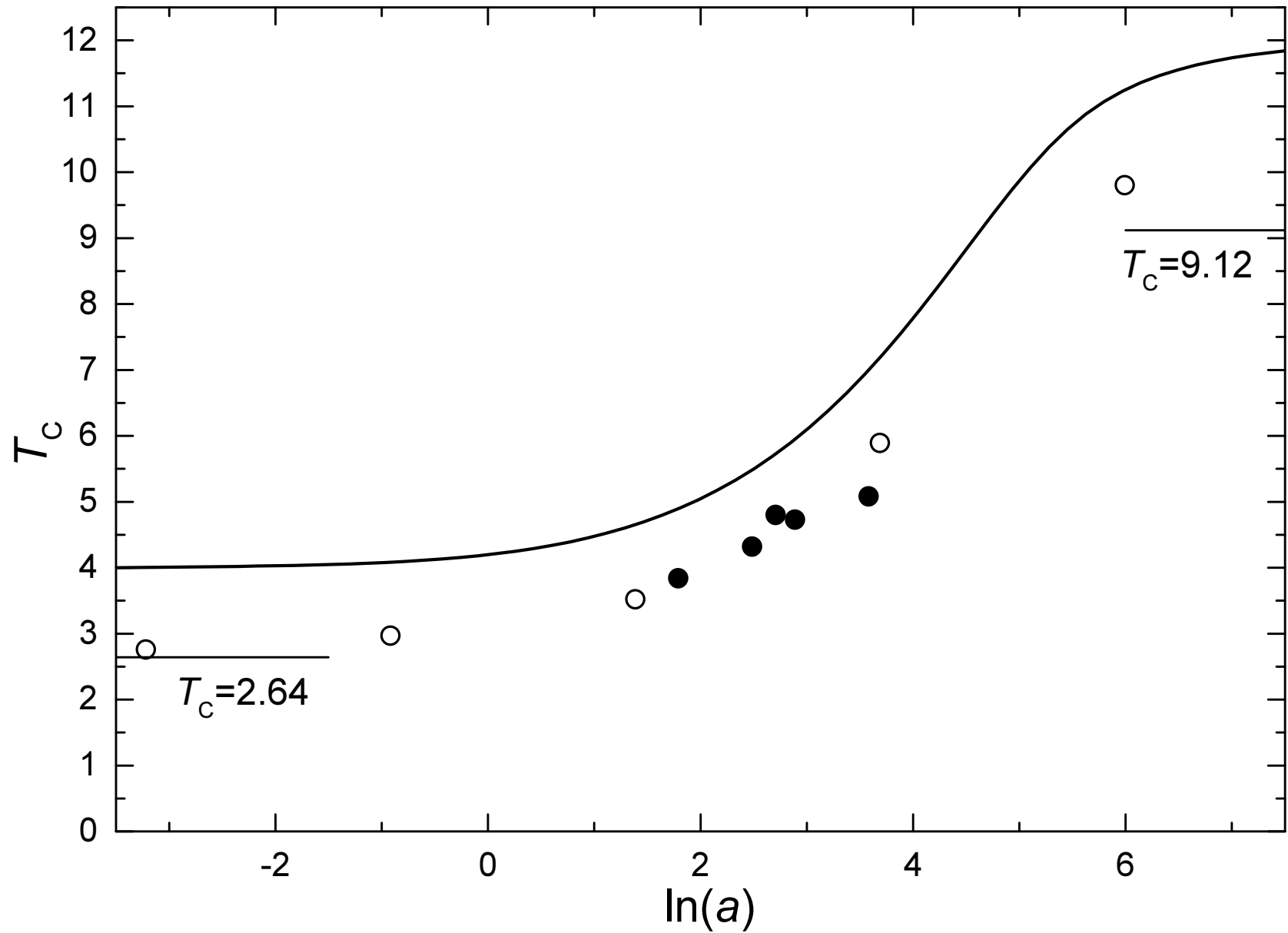
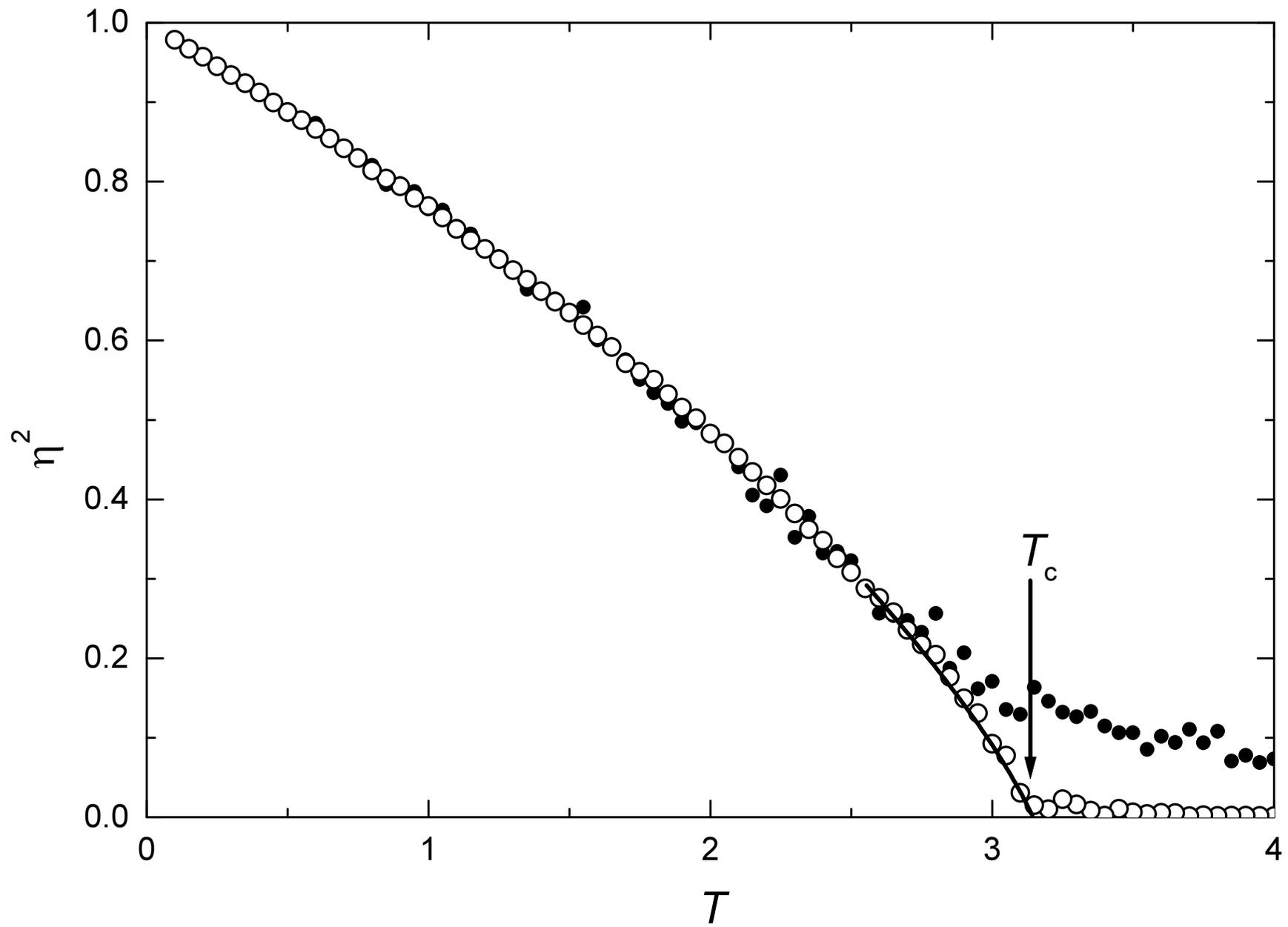
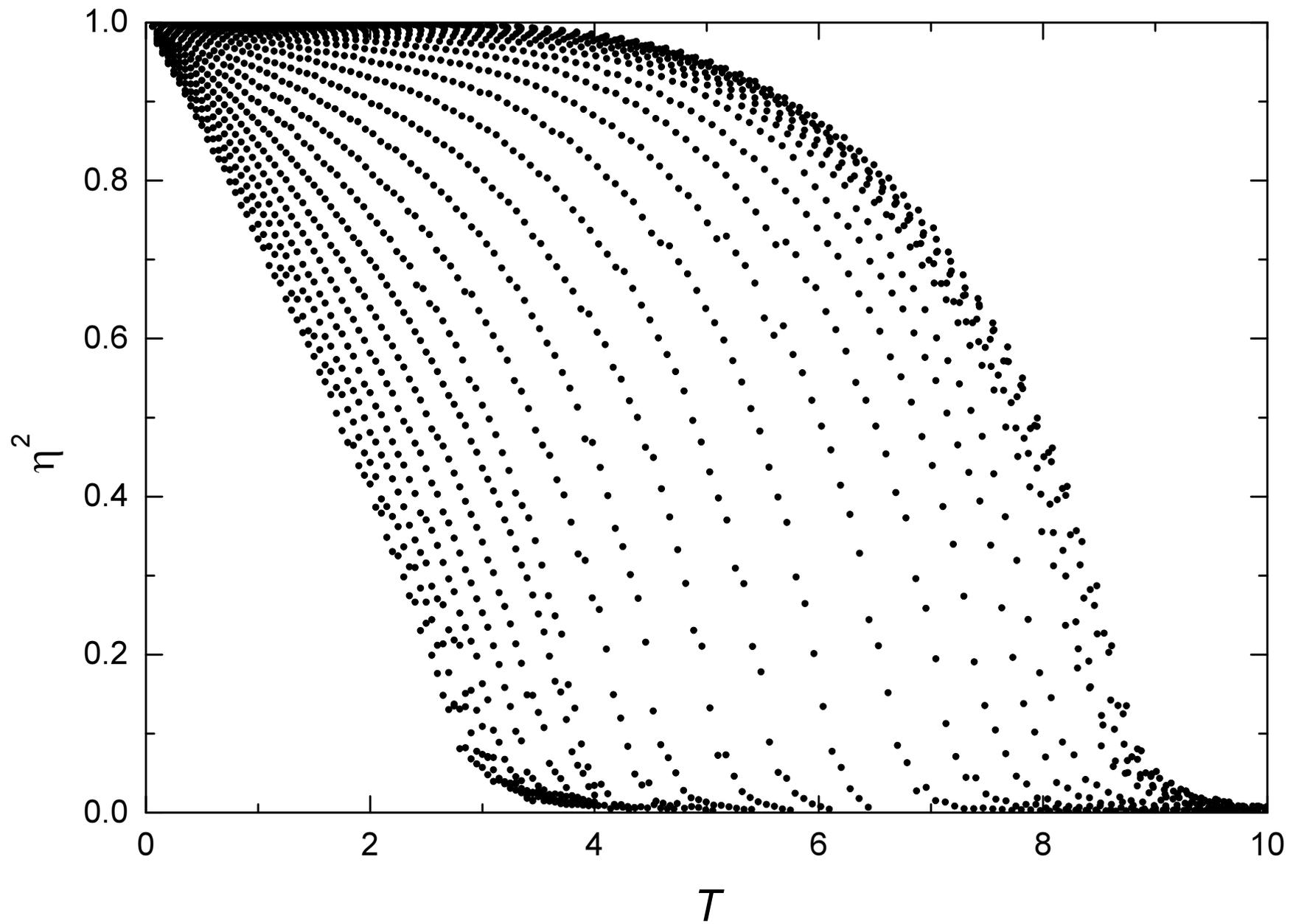
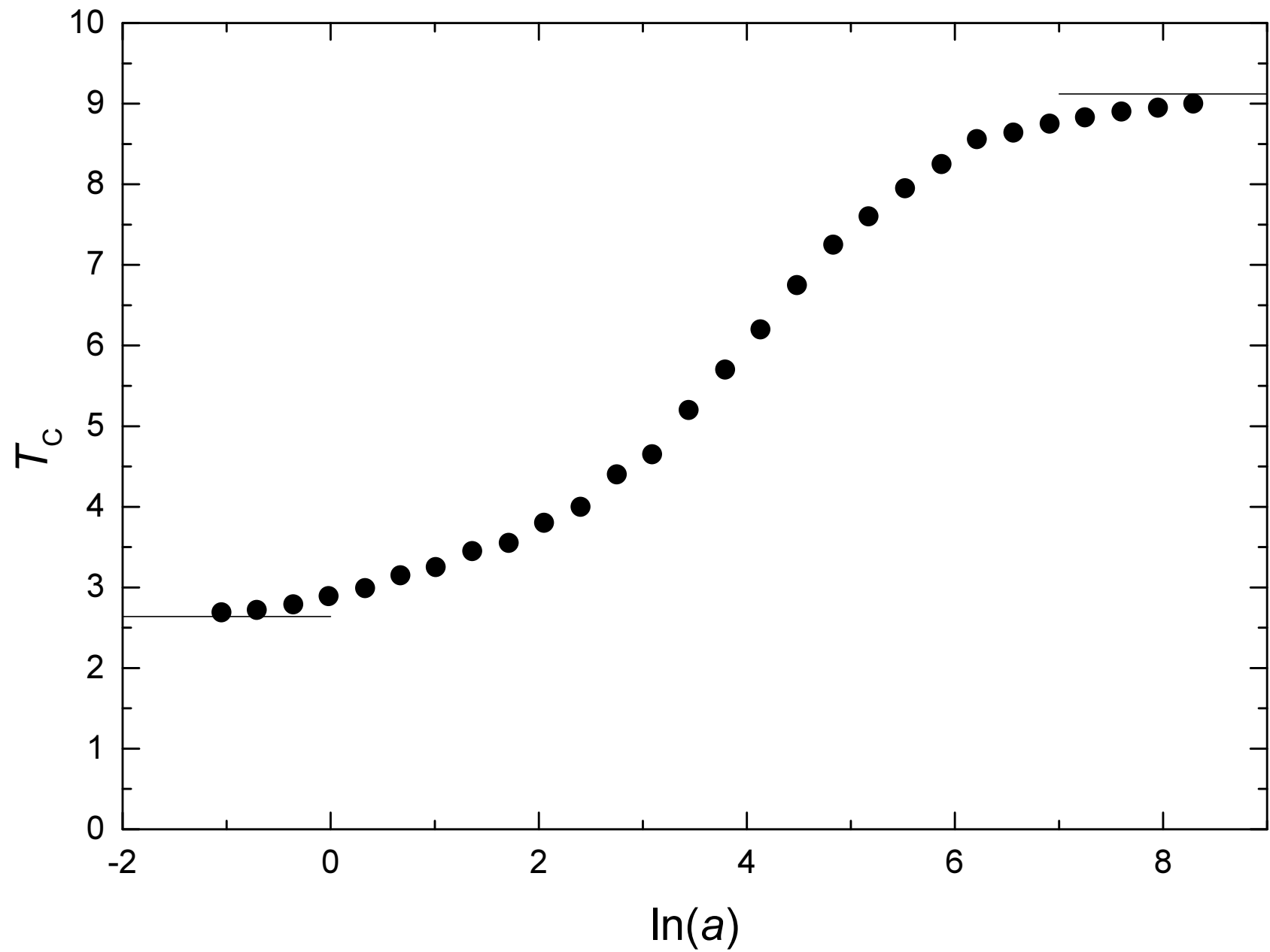
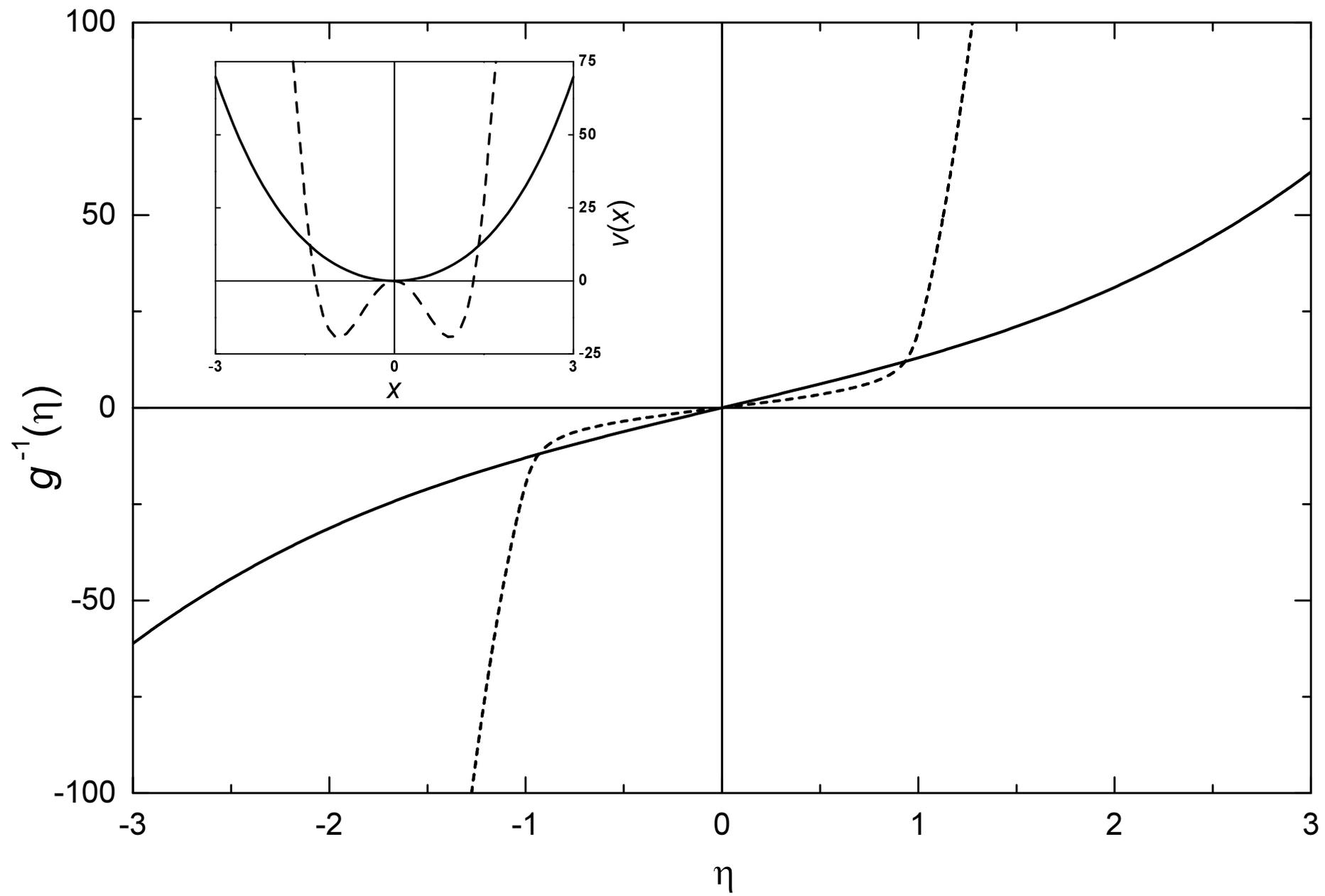
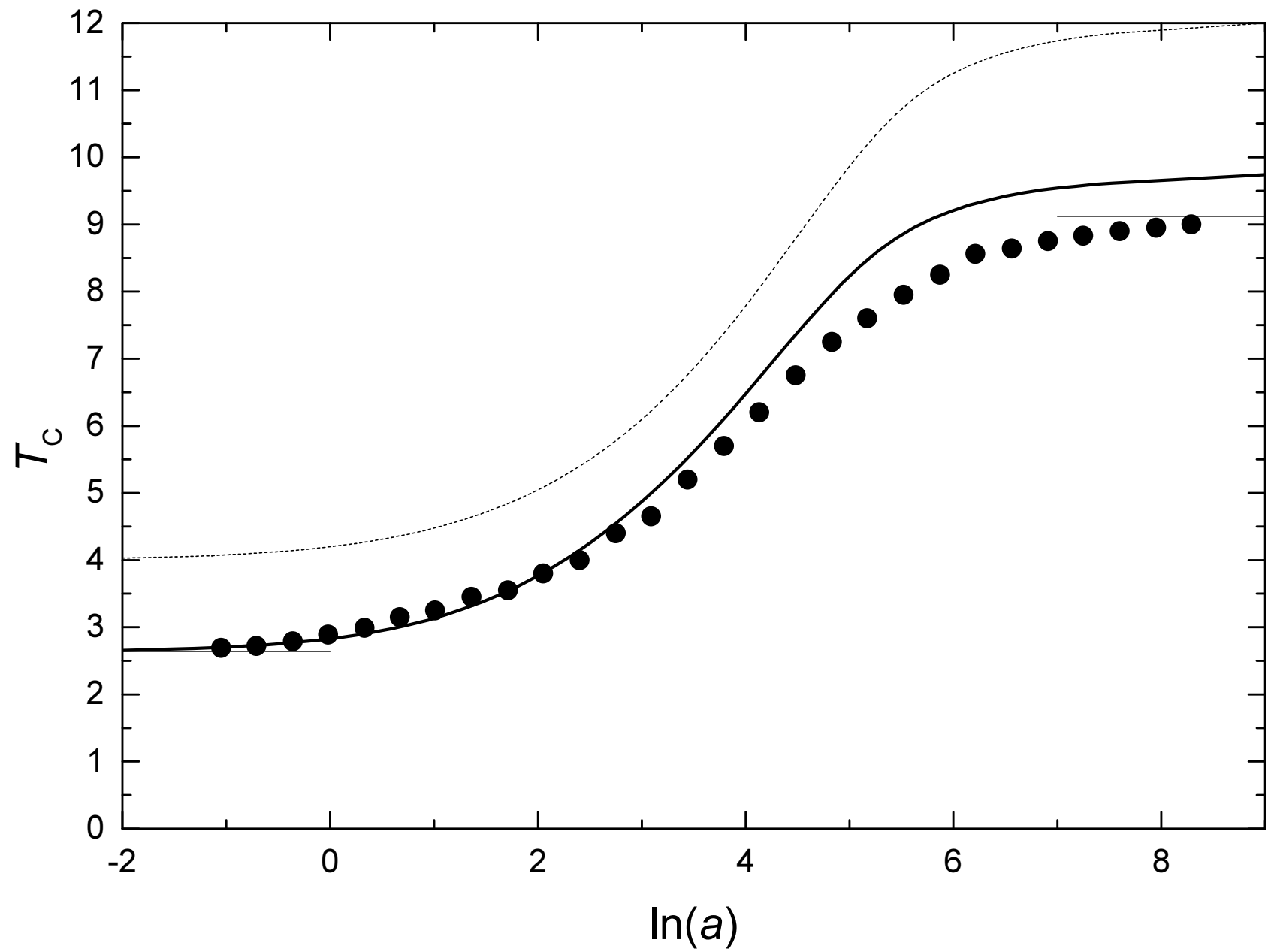
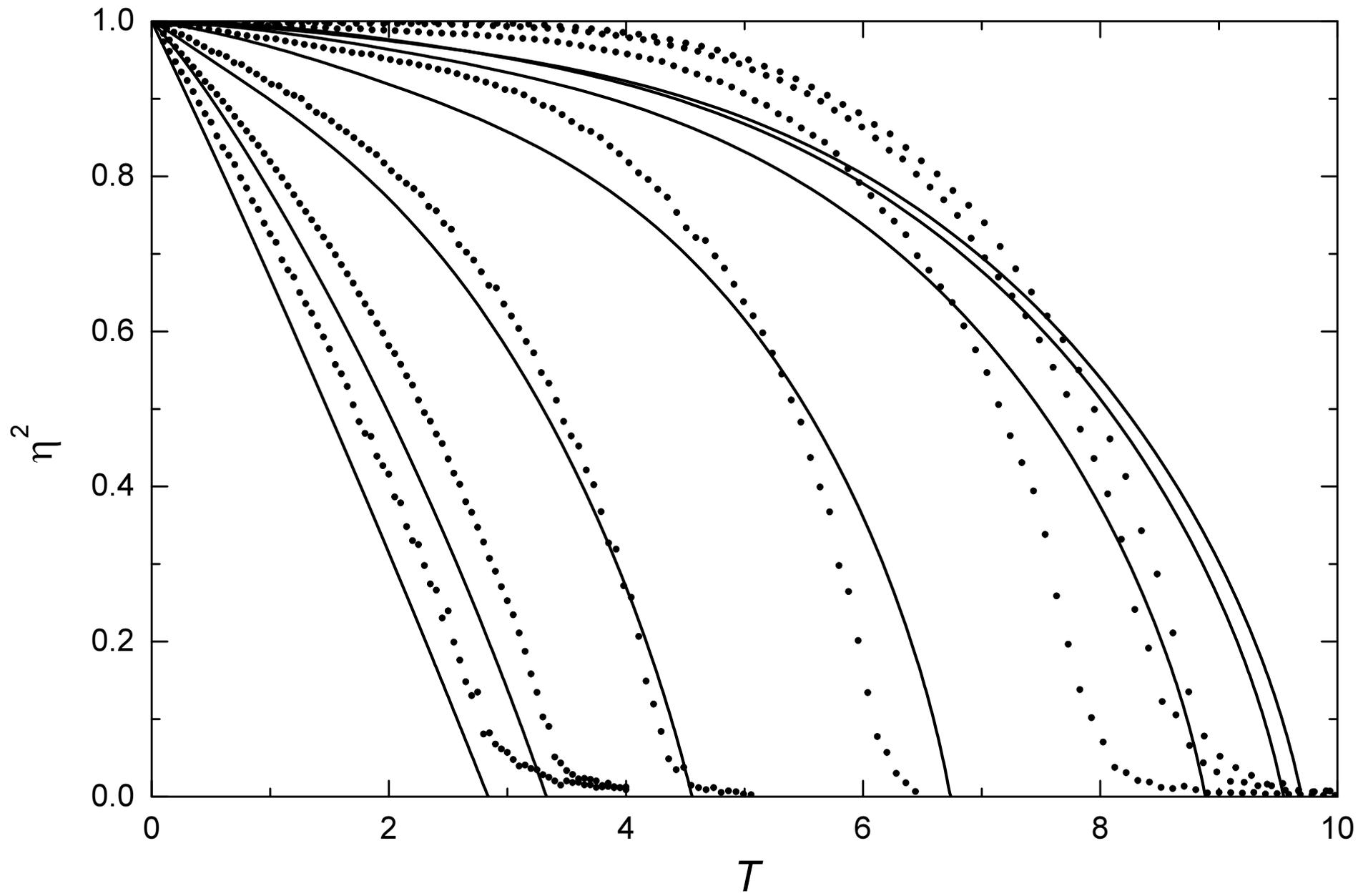
\end{document}